\documentclass[useAMS,usenatbib]{mn2e}

\usepackage{epsfig}
\usepackage{apjfonts}
\usepackage{mathptmx}

\newcommand{\sbeamp}[5]{{$#1\mbox{$''\mskip-7.6mu.\,$}#2$} $\times$ {$#3\mbox{$''\mskip-7.6mu.\,$}#4$}; $+#5^\circ$}
\newcommand{\sbeamm}[5]{{$#1\mbox{$''\mskip-7.6mu.\,$}#2$} $\times$ {$#3\mbox{$''\mskip-7.6mu.\,$}#4$}; $-#5^\circ$}

\newcommand{\apj}{ApJ}
\newcommand{\araa}{ARAA}
\newcommand{\aap}{A\&A}
\newcommand{\apjl}{ApJ}
\newcommand{\aj}{AJ}
\newcommand{\mnras}{MNRAS}

\title{ALMA and VLA observations of the outflows in IRAS 16293--2422}

\author[Laurent Loinard et al.]{Laurent Loinard$^{1,2}$, Luis A.\ Zapata$^1$, Luis F.\ Rodr\'{\i}guez$^1$, Gerardo Pech$^1$, Claire J.\ Chandler$^3$, 
\newauthor Crystal L., Brogan$^4$,  David J.\ Wilner$^5$, Paul T.P.\ Ho$^{5,6}$, B\'ereng\`ere Parise$^2$, Lee W.\ Hartmann$^7$, 
\newauthor Zhaohuan Zhu$^8$, Satoko Takahashi$^6$, and Alfonso Trejo$^6$\\
$^{1}${Centro de Radiostronom\'{\i}a y Astrof\'{\i}sica, Universidad Nacional Aut\'onoma de M\'exico, 58089 Morelia, Michoac\'an, M\'exico}\\
$^{2}${Max-Planck-Institut f\"ur Radioastronomie, Auf dem H\"ugel 69, 53121 Bonn, Germany}\\
$^{3}${National Radio Astronomy Observatory, P.O. Box O, Socorro, NM 87801}\\
$^{4}${National Radio Astronomy Observatory, 520 Edgemont Road, Charlottesville, VA 22903-2475}\\
$^{5}${Harvard-Smithsonian Center for Astrophysics, 60 Garden Street, Cambridge, MA 02138}\\
$^{6}${Academia Sinica Institute of Astronomy and Astrophysics, Taipei, Taiwan}\\
$^{7}${Department of Astronomy, University of Michigan, 500 Church St., Ann Arbor, MI 48109, USA}\\
$^{8}${Department of Astrophysical Sciences, 4 Ivy Lane, Peyton Hall, Princeton University, Princeton, NJ 08544}}

\begin{document}

\date{Accepted \today. Received \today; in original form \today}

\pagerange{\pageref{firstpage}--\pageref{lastpage}} \pubyear{2012}

\maketitle

\label{firstpage}

%\date{Received September 15, 1996; accepted March 16, 1997}

\begin{abstract}
We present ALMA and VLA observations of the molecular and ionized gas at 0.1-0.3$''$ resolution in the Class 0 protostellar
system IRAS 16293-2422. These data clarify the origins of the protostellar outflows from the deeply embedded sources in this 
complex region. Source A2 is confirmed to be at the origin of the well known large scale north-east--south-west flow. The 
most recent VLA observations reveal a new ejection from that protostar, demonstrating that it drives an episodic jet. The 
central compact part of the other known large scale flow in the system, oriented roughly east-west, is well delineated by 
the CO(6-5) emission imaged with ALMA and is confirmed to be driven from within component A. Finally, a one-sided 
blueshifted bubble-like outflow structure is detected here for the first time from source B to the north-west of the system. 
Its very short dynamical timescale ($\sim$ 200 yr), low velocity, and moderate collimation support the idea that source 
B is the youngest object in the system, and possibly one of the youngest protostars known. 
\end{abstract}

\section{Introduction}

Outflows constitute one of the most spectacular, emblematic, and relevant early manifestations of 
the star formation process (e.g.\ Lada 1985; Arce et al.\ 2007). By removing angular momentum, 
they allow accretion to proceed (Bodenheimer 1995), and through their injection of mechanical 
energy, they can affect the overall evolution of star-forming regions (Reipurth et al.\ 1997). Their 
morphologies can also provide clues on the properties of the driving sources. For instance, the 
degree of collimation of outflows is  believed to correlate with the age and mass of the driving 
protostars (Machida et al.\ 2008; Arce et al.\ 2007), while multipolar outflows are often an 
indication that the central engine is a multiple system (e.g.\ Carrasco-Gonz\'alez et al.\ 2008). 

IRAS~16293--2422 (Ceccarelli et al.\ 2000; 
Crimier et al. 2010) is located at a distance of 120 pc (Loinard et al.\ 2008) in the eastern part 
of the Ophiuchus complex. Soon after its discovery (Walker et al.\ 1986), it was found to power 
two bipolar outflow structures (Mizuno et al.\ 1990). One is oriented in the north-east--south-west 
direction (at a position angle of about 55$^\circ$ and hereafter called the NE-SW flow), 
while the other is oriented nearly east-west (P.A.\ $\sim$ 110$^{\circ}$, hereafter the EW flow). 
The existence of at least two protostellar sources was also established early thanks to high 
resolution radio observations (Wootten 1989, Mundy et al.\ 1992). These objects are known 
as component A to the south-east, and component B to the north-west, and their projected 
separation is about 5$''$ (Figure 1). While component B remains single even at the highest 
angular resolution available ($\sim$ 0.05 arcsec; Rodr\'{\i}guez et al.\ 2005; Chandler et al.\ 
2005), component A breaks up into two centimeter sub-components called A1 and A2 at resolutions 
better than about 0.2$''$ (Figure 1). Because the 
relative orientation of the A1/A2 pair at the time of its discovery was very similar to the direction 
of the NE-SW flow, A1 was initially believed to be an ejecta from A2. However, analysis of the 
relative motion of A1 and A2 favors a scenario where these two sources trace the two stars in a 
binary system (Loinard 2002; Chandler et al.\ 2005; Loinard et 
al.\ 2007, Pech et al.\ 2010). Sub-arcsecond submillimeter continuum observations obtained
with the SMA (Submillimeter Array) revealed additional structure in component A (Chandler 
et al.\ 2005). While the emission is dominated by a bright compact source (called Aa) centered 
almost exactly between A1 and A2, an additional source (Ab) was detected to the north-east of 
component A (Figure 1). This source has not been detected at any other wavelengths so far, and 
its exact nature remains mysterious.

The NE-SW outflow has long been known to originate in component A (e.g.\ Hirano et al.\ 
2001; Castets et al.\ 2001). This association was unambiguously confirmed, and was further
constrained, when Loinard et al.\ (2007) witnessed, at radio wavelengths, the ejection from 
A2 of a bipolar system of ejecta which, subsequently, moved away from A2 roughly along 
the direction of the NE-SW flow (Pech et al.\ 2010). The EW flow was recently shown to also 
originate from within component A by Yeh et  al.\ (2008) who presented SMA CO(2-1) and 
CO(3-2) observations 
at an angular resolution of a few arcseconds. These SMA data, however, 
have insufficient angular resolution to distinguish between A1 and A2 as the driving source 
for the EW flow. Interestingly, there is little evidence for outflow activity from component B. 
J{\o}rgensen et al.\ (2011) have presented submillimeter (SMA) observations of IRAS~16293--2422 
that resolve the A and B components and do not show strong 
indications for high velocity gas toward B. Yeh et al.\ (2008) reported the detection of a 
compact blue-shifted CO structure to the south-east of source B (their structure b2), but their 
data was insufficient to decide whether or not this was a compact flow from source B. 
Rao et al.\ (2009) also detected this structure,  and argued that it was part of an additional 
outflow driven by source A and oriented at a position angle of about 145$^\circ$.
In this {\it Letter}, we present combined ALMA (Atacama Large Millimeter/submillimeter Array) 
and VLA (Karl G.\ Jansky Very Large Array) observations of IRAS~16293--2422 that help elucidate 
the origin and nature of the outflows in this complex young stellar system.

\section{Observations}

\subsection{ALMA data}

IRAS~16293--2422 was observed at $\lambda$ = 0.45 mm with fifteen 12-m antennas of ALMA on 
April 2012, during the science verification data program. The 105 independent baselines ranged in 
projected length from 26 to 403 m. The primary beam of ALMA at 0.45 mm has a FWHM 
of 8$''$, so the observations were made in mosaicking mode with half-power point spacing 
between field centers to cover both source A and source B. The digital correlator 
was configured in 4 spectral windows of 1875 MHz and 3840 channels each.  This provides 
a channel spacing of 0.488 MHz ($\sim$ 0.2 km s$^{-1}$) per channel, but the spectral 
resolution is a factor of two lower (0.4 km s$^{-1}$) due to online Hanning smoothing.  Observations of Juno 
provided the absolute scale for the flux density calibration while observations of the quasars 
J1625$-$254 and NRAO530 (with flux densities of 0.4 and 0.6 Jy, respectively) provide the gain 
calibration. The quasars 3C279 and J1924-292 were used for the bandpass calibration.
The data were calibrated, imaged, and analyzed using the Common Astronomy Software Applications 
(CASA) and KARMA software (Gooch 1996).  The continuum
emission as well as many spectral lines were detected in these ALMA data (see J{\o}rgensen et al.\
2012). Here we concentrate on the analysis
of the continuum emission and of the CO(J=6-5; $\nu$=0) at a rest frequency of 691.47308 GHz.
The r.m.s.\ noise, measured in line-free channels on either side of the CO line, is about 50 mJy beam$^{-1}$ 
at an angular resolution of \sbeamm{0}{32}{0}{18}{69}. Only a small number of line-free channels 
were averaged to calculate the continuum, so the r.m.s.\ noise of the continuum image is only 
moderately better (20 mJy beam$^{-1}$) than that in the individual spectral channels. 
%Although
%a much deeper continuum image could be formed, our analysis is not sensitivity-limited and will 
%not be affected.

\subsection{VLA data}

IRAS~16293--2422 was observed with the VLA in its most extended (A) configuration on 2011, June 
5 and 8 at $\lambda$ = 7 mm, and on 2011, August 12 at $\lambda$ = 4 cm. In all cases, the absolute 
flux calibrator was 3C286 (J1331$+$3030), while the amplitude and phase gains were monitored using 
observations of J1625$-$2527. At 7 mm, sixteen contiguous spectral windows, each 
containing 64 spectral channels 2 MHz-wide were recorded simultaneously to produce a total bandwidth 
of 2 GHz (between 40 and 42 GHz). At  4 cm, eight contiguous spectral windows, each 
containing 64 spectral channels 2 MHz-wide were recorded simultaneously to produce a total bandwidth 
of 1 GHz (between 6.8 and 7.8 GHz).\footnote{A second spectral band, corresponding to the frequency 
range from 4.2 to 5.2 GHz was observed simultaneously with the 4 cm data presented here, but will 
not be discussed in this {\it Letter}.} 
The data calibration was performed using the CASA software. The data 
were manually inspected and flagged, and calibrated following standard 
procedures. The calibrated visibilities at 4 cm were imaged using a nearly uniform weighting 
scheme (Robust parameter set to $-$2 in CASA) yielding a synthesized beam of \sbeamm{0}{44}{0}{16}{11}
and a noise level of 16 $\mu$Jy beam$^{-1}$. The visibilities at 7 mm were imaged both with 
natural weighting (Robust = $+$2) to obtain the highest sensitivity (at the detriment of angular resolution), 
and with intermediate weighting (Robust = 0) to obtain higher angular resolution (at the detriment of sensitivity). 
The synthesized beam was \sbeamp{0}{14}{0}{12}{69} and the sensitivity was 19 $\mu$Jy beam$^{-1}$ 
in the former case, while they were \sbeamm{0}{08}{0}{07}{35} and 31 $\mu$Jy beam$^{-1}$ in the latter. 
While the sensitivity at 4 cm is near the theoretically expected value given the integration time,
the sensitivity at 7 mm is about 50\% worse than expected. Additional analysis will be needed to understand
the origin of this degradation, but we note that in spite of this effect, the maps presented here are roughly 
5 times deeper that the best previous 7 mm images (Rodr\'{\i}guez et al.\ 2005).

\begin{figure}
  \centerline{\includegraphics[height=0.45\textwidth,angle=-90]{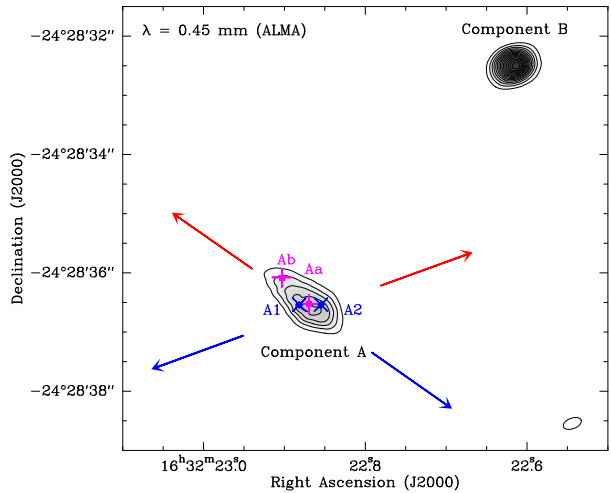}}
  \caption{Sub-millimeter continuum image of the IRAS 16293--2422 system 
  from ALMA. The contours run from 0.2 to 4 Jy beam$^{-1}$ by steps of 0.2 Jy 
  beam$^{-1}$; the synthesized beam (\sbeamm{0}{32}{0}{18}{69}) is shown at the 
  bottom-right. The noise level is 0.02 mJy beam$^{-1}$.
  The two main components (A and B) are labelled, and the direction of the
  two outflows driven from component A are indicated (from Mizuno et al.\ 1990). 
  The sub-millimeter peaks Aa and Ab from Chandler et al.\ (2005) and the centimeter 
  sources A1 and A2 are shown.}
    \label{fig:almacont}
\end{figure}

\begin{figure}
  \centerline{\includegraphics[width=0.45\textwidth,angle=0]{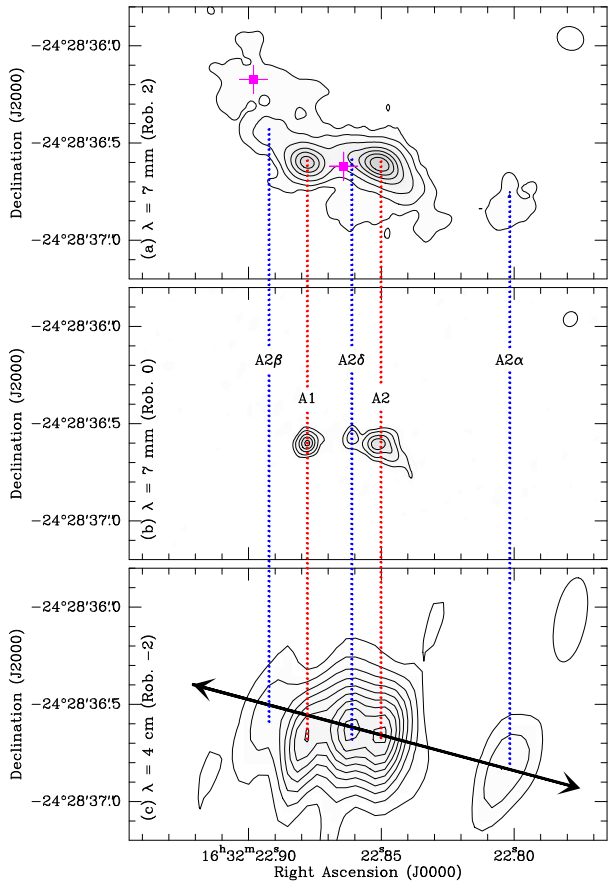}}
  \caption{VLA continuum images of IRAS 16293--2422. (a) 7 mm image
  reconstructed with natural weighting to optimize sensitivity. The synthesized
  beam (\sbeamp{0}{14}{0}{12}{69}) is shown at the top-right of the panel, and 
  the sensitivity is 19 $\mu$Jy beam$^{-1}$. The contours are at 70, 140, 300, 600, 
  900, etc. $\mu$Jy beam$^{-1}$. For reference, the position of the sub-millimeter 
  peaks Aa and Ab are shown as magenta crosses. Note the faint extension to the 
  NE coincident with Ab and the extension of the sub-millimeter ALMA image, and 
  the faint emission associated with the ejecta A2$\alpha$ and A2$\beta$. (b)  
  7 mm image reconstructed with intermediate weighting. The synthesized
  beam (\sbeamm{0}{08}{0}{07}{35}) is shown at the top-right of the panel, and 
  the sensitivity is 31 $\mu$Jy beam$^{-1}$. The contours are at 150, 300, 600, 
  900, etc. $\mu$Jy beam$^{-1}$. Note the new ejecta A2$\delta$. (c) 4 cm image 
  where all the sources and ejecta are seen. The synthesized beam 
  (\sbeamm{0}{44}{0}{16}{11}) is shown at the top-right of the panel, and the sensitivity 
  is 16 $\mu$Jy beam$^{-1}$. The first contour and the contour spacing are at $\mu$Jy 
  beam$^{-1}$.  The double arrow shows the direction (at P.A.\ 70$^\circ$) along which the ejecta
  A2$\alpha$, A2$\beta$, and A2$\delta$ were launched.}
      \label{fig:jvla}
\end{figure}

\section{Results and discussion}

\subsection{Continuum emission and the NE-SW flow}

The sub-millimeter continuum emission ($\lambda$ = 0.45 mm) measured by ALMA reveals the usual 
double source structure of IRAS~16293--2422 (Figure \ref{fig:almacont}). Component B is resolved but 
compact, with a peak brightness temperature of about 160 K. The total flux measured here with ALMA
($\sim$ 12.2 Jy) is consistent with the trend observed at lower frequencies that the spectral index of
source B is between 2 and 2.5 from centimeter to submillimeter wavelengths (Chandler et al.\ 2005).
Component A is elongated along the NE-SW direction, with its peak at the position of source Aa, and 
the faint extension to the NE corresponding to the position of source Ab reported by Chandler et al.\ 
(2005). Interestingly, the centimeter sub-components A1 and A2 are not identifiable in this image, 
although the angular resolution would be sufficient to separate them. The peak brightness temperature 
of the sub-millimeter emission in component A is about 30 K. The overall properties of the ALMA image
of component A suggest that the sub-millimeter continuum is dominated by the circumbinary envelope 
rather than by the individual disks around the protostars. This is not particularly surprising since the 
individual circumstellar disks are expected to be heavily tidally truncated in this very compact binary system
(projected separation of order 40 AU). In contrast, the continuum at 7 mm is dominated by free-free 
radiation from the bases of the thermal jets and enables us to pin-point the embedded sources A1 and 
A2 (Figure  \ref{fig:jvla}). This demonstrates the complementarity of ALMA and the VLA to study young 
stellar sources. 

When it is reconstructed to optimize sensitivity, the 7 mm emission from 
source A also reveals a faint elongation toward the NE similar to that seen in the sub-millimeter map, and 
extended emission surrounding the sources A1 and A2. Like the sub-millimeter continuum, this
extended emission likely traces the circumbinary envelope. The emission associated with the extension
of source A to the north-east (i.e.\ with source Ab from Chandler et al.\ 2005) contributes roughly 1.4 mJy 
at 41 GHz, 0.5 Jy at 305 GHz (Chandler et al.\ 2005) and 3 Jy at 696 GHz (all values with relative uncertainties 
of order 50\%). This yields a spectral index of 2.7 $\pm$ 0.5, consistent with that of the more compact 
emission from source A (Chandler et al.\ 2005). We argue that this elongation traces faint extended emission 
from the circumbinary envelope (possibly related to dust accumulated in that direction by the NE-SW flow), 
rather than a new protostellar source in the system.

Both the 7 mm and the 4 cm continuum maps (Figure \ref{fig:jvla}) reveal a new compact source 
to the NE of source A2. This is strikingly similar to the situation encountered in the 2006 image of 
IRAS~16293--2422 at $\lambda$ = 1.3 cm (Loinard et al.\ 2007) when a bipolar ejection from A2 
was seen for the first time. The bipolar ejection that occurred in 2006 produced the sources 
A2$\alpha$ and A2$\beta$ that are still clearly seen in the 4 cm image (Figure \ref{fig:jvla}c) and, faintly, 
in the 7 mm map (Figure \ref{fig:jvla}a). They are moving away from A2 at projected velocities 
of several tens of km s$^{-1}$ (Pech et al.\ 2010). We interpret the new source to the NE of A2 (which 
we shall call A2$\delta$) as a new ejection from A2. In the high resolution 7 mm image, there is a 
faint extension to the SW of A2 which might correspond to the bipolar counterpart of A2$\delta$. We 
note that in the first image following the ejection of the A2$\alpha$/A2$\beta$ pair (Loinard et al.\ 2007), 
the SW ejecta (A2$\alpha$) was not immediately detached from A2 either.

\begin{figure}
  \centerline{\includegraphics[width=0.43\textwidth,angle=0]{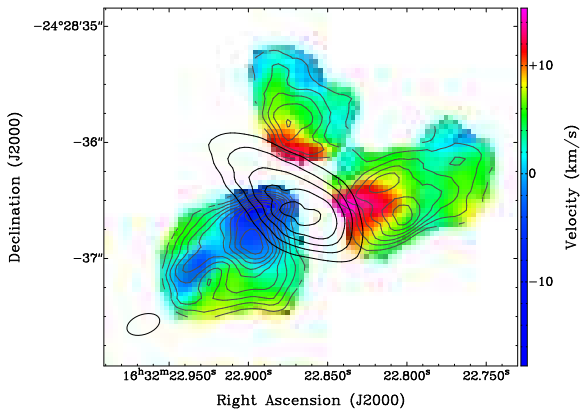}}
  \caption{Integrated intensity of the weighted velocity (moment 1) color map
of the CO(6-5) emission from source A overlaid in contours with the
0.45 mm continuum emission (black thick line) and the integrated
intensity line emission (moment 0) of the CO(6-5) (grey thin line).
The black contours are the same as in Figure \ref{fig:almacont} (0.2 to 1.2 Jy beam$^{-1}$ in
steps of 0.2 Jy beam$^{-1}$). The grey contours are from 10\% to 90\% with
steps of 10\% of the peak of the CO line emission; the peak is 3.2 $\times$ 10$^4$ Jy Beam$^{-1}$ km s$^{-1}$.
The color-scale bar on the right indicate the LSR velocities in km s$^{-1}$. 
The synthesized beam of the continuum image is shown in the bottom left corner of the image. }
 \label{fig:outflowA}
\end{figure}

IRAS~16293--2422 has been regularly monitored at centimeter wavelengths since the late 1980s 
(Chandler et al.\ 2005) and the 2006 ejection was the first to be detected. The new ejection reported
here suggests that source A2 might be entering a period of enhanced outflow activity. Regular
observations in the coming years and decades will be required to confirm this possibility, and will 
enable a detailed study of the kinematics and behavior of the ejecta. Interestingly, A2$\alpha$, 
A2$\beta$, A2$\delta$, and A2 itself are well aligned (Figure \ref{fig:jvla}c) but the position
angle ($\sim$ 70$^\circ$) of the line joining them is somewhat larger than the position angle ($\sim$ 
55$^\circ$) of the large-scale outflow known to originate from A2 (e.g.\ Figure 5 in Mizuno et al.\ 1990).
This suggests that the jet driven by source A2 is precessing as a consequence of the A1-A2 binarity.
We note, finally, that there is no detectable CO(6-5) emission associated with the NE-SW flow
in the ALMA observations (Figure \ref{fig:outflowA}). This is in agreement with the results of Yeh et al.\ 
(2008) who also failed to detect such emission in their SMA CO observations at arcsecond resolution.

\subsection{The EW flow}

Strong CO(6-5) is detected in the ALMA data around the position of source A. In agreement with the
results of Yeh et al.\ (2008), it is largely concentrated to an EW structure with blueshifted emission to 
the east and redshifted emission to the west;  This emission traces the central part of the well-known
EW (PA $\sim$ 110$^\circ$) flow (Figure \ref{fig:outflowA}). In spite of the high angular resolution of the 
ALMA observations, the origin of the flow stall cannot be traced back to a specific source within component 
A.

\subsection{A compact outflow from source B}

\begin{figure}
  \centerline{\includegraphics[width=0.43\textwidth,angle=0]{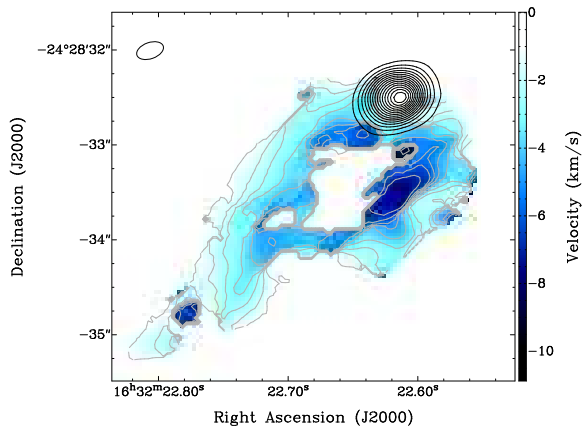}}
  \caption{Integrated intensity of the weighted velocity (moment 1) color map
of the CO(6-5) emission from source B overlaid in contours with the
0.45 mm continuum emission (black thick line) and the velocity scale of the CO(6-5) (grey thin line).
The black contours are the same as in Figure \ref{fig:almacont} (starting at and in step of 0.2
Jy beam$^{-1}$). The color-scale bar on the right indicate the LSR velocities in 
km s$^{-1}$. The synthesized beam of the continuum image is shown in the upper left corner of the image. }
    \label{fig:outflowB}
\end{figure}

As reported by Yeh et al.\ (2008) and confirmed by Rao et al.\ (2009), there is very strong blue-shifted CO 
emission to the south of source B (the location of component b2 in Yeh et al.\ 2008). Figure \ref{fig:outflowB} 
displays the first moment CO(6-5) map of this component (in colors) overlaid with the 0.45 mm continuum 
image (in contours). It is clear that the CO emission defines a 3$''$-long bubble-like structure oriented along 
the south-east--north-west direction (at P.A.\ $\sim$ 130$^\circ$) with source B at its north-west apex. This 
structure points from source B toward source A, but there appears to be no material connection (i.e.\ no 
bridge of emission) between the two. In particular, its south-east apex (corresponding to 
the point nearest to A) is located about 2$''$ to the north-west of source A. We conclude that this structure is 
associated with source B, and unrelated to source A. We should point out, here, that there is a significant
amount of extended CO emission around the systemic velocity of IRAS~16293--2422 (v$_{lsr}$ $\sim$ 4 km 
s$^{-1}$) in the ALMA observations. This emission is poorly recovered by the interferometer, and its structure 
cannot be assessed; we filtered it out by ignoring the velocity channels around 4 km s$^{-1}$, but note that
our fluxes (as given, for instance, in Figure \ref{fig:outflowB}) are clearly underestimated since they do not 
include the ambient gas. Interestingly, Rao et al.\ (2009) recently reported on the 
possible detection of a new outflow in IRAS~16293--2422 seen in SiO(8-7) along a position angle very similar 
to that of the CO structure seen here. The structure seen in SiO, however, is most prominent south of source
A, and entirely at positive velocities. While there is SiO(8-7) emission in the general direction of the CO structure
reported here, it is at $+$4 km s$^{-1}$ rather than at --4 km s$^{-1}$. Higher resolution SiO observations will 
clearly be needed to establish the relation (if there is any) between the SiO structure reported by Rao et al.\ 
(2009) and the CO structure reported here.

To our knowledge, ours is the first direct indication for the existence of an outflow driven by source B. Interestingly, 
there is no strong redshifted counterpart to the north-west.
While this might reflect an intrinsic asymmetry of the flow, it could also at least partly reflect the structure of the 
circumstellar environment. As discussed by Pineda et al.\ (2012) and Zapata et al.\ (in prep.), at sub-millimeter 
wavelengths, both the CO lines and the continuum emission toward source B are optically thick, so the red-shifted 
emission might be at least partly hidden from view. Alternatively, IRAS~16293--2422 might be located close to the 
back side of the cloud where it is located. Besides explaining the lack of a red-shifted lobe, this would be consistent 
with the very high optical depth toward source B. 

Since the systemic velocity of source B is about 3 km s$^{-1}$ (J{\o}rgensen et al.\ 2011) while the most negative 
velocities reached by the outflow are $\sim$ $-$8 km s$^{-1}$, the (radial) expansion velocity of the bubble relative 
to source B is about 10 km s$^{-1}$. At the distance of Ophiuchus, the angular extent of the bubble (3$''$) 
corresponds to 5.4 $\times$ 10$^{15}$ cm (360 AU), and the dynamical age of the flow is of order 200 years. 
While this calculation might only provide a crude estimate of the true age of the structure, similar calculations 
for the other outflows in the system (Mizuno et al.\ 1990) yield much larger dynamical ages, of order 10$^4$ 
years. This is consistent with the commonly held view that source B is the youngest object in IRAS~16293--2422 
(e.g.\ Chandler et al.\ 2005). It could be argued that the true angular extent of the outflow is significantly larger than 
reported here, because the CO(6-5) line might only pick the most highly excited gas in the vicinity of source B. This 
is unlikely because lower J CO observations (e.g.\ Yeh et al.\ 2008) did not reveal any conspicuous emission on larger 
scales along the direction of this structure. 

The 7 mm continuum emission around source B has usually been interpreted as a nearly face-on disk (Rodr\'{\i}guez  
et al.\ 2005). Recent ALMA spectroscopic observations further revealed unambiguous evidence for infall and rotation 
(Pineda et al.\ 2012, Zapata et al.\ in prep.) toward that source, and the rotation is also consistent with a nearly face-on 
orientation. In this configuration, the flow driven by source B ought to be oriented nearly along the line of sight, and the 
expansion velocity would be mostly radial. This suggests that the true expansion velocity of the flow is not significantly 
larger than the observed 10 km s$^{-1}$. 

The overall morphology of the outflow driven by source B is qualitatively different from that of typical flows powered
by low-mass young stars. In particular, there is no evidence for a collimated jet-like structure along the symmetry axis. 
Instead, the highest (i.e.\ most negative) velocities are associated with the inner layer of the structure, while the outer 
layers are at less negative velocities (typically at about $-$2 km s$^{-1}$; see Figure \ref{fig:outflowB}). This behavior 
suggests that source B drives a moderately collimated wind, that impinges into the stationary circumstellar medium, 
in a situation resembling that described theoretically by Shu et al.\ (1991).

%The structure of the outflow driven by source B is also strongly reminiscent of the outburst suffered by the somewhat 
%more evolved young star XZ Tauri in the 1980s and monitored using HST observations over the following few 
%decades (Krist et al.\ 2008). The similarities between the two structures include the moderate collimation of the flow, 
%and their strong asymmetries (the outburst suffered by XZ Tau is, like the flow reported here, much more prominent 
%on one side than the other). Even the linear sizes (of order 500 AU) are similar in both cases. It is noteworthy, however, 
%that the dynamical age of the flow reported here is about 10 times larger than that of the outflow in XZ Tau. This reflects 
%the difference in the expansion velocity of the two outflows ($\sim$ 10 km s$^{-1}$ in IRAS~16293--2422B against 
%$\sim$ 100 km s$^{-1}$ in XZ Tau). 

In summary, the flow driven by source B has a number of peculiar properties. It is fairly slow ($\sim$ 10 km s$^{-1}$),
outburst-like, and only moderately collimated. Those are characteristics theoretically expected from flow driven by 
adiabatic (first) cores (Machida et al.\ 2008, Price et al.\ 2012; see also Chen et al.\ 2010, Enoch et al.\ 2010, Dunham 
et al.\ 2011, Pineda et al.\ 2011 for observational aspects). However, the evidence for strong accretion reported by 
Pineda et al.\ (2012) and Zapata et al.\ (in prep.) and the fact that component B is such a bright submillimeter continuum
source would be more consistent with the idea that it has already entered the protostellar phase. In any case, the 
properties of the outflow driven by this object do suggest that it might be one of the youngest known protostars. 

\section{Conclusions and perspectives}

In this {\it Letter}, we used combined ALMA and VLA observations to examine the outflow activity in the very young
(Class 0) protostellar system IRAS~16293--2422. The well-known NE-SW and E-W large-scale molecular flows are 
confirmed to be driven from within component A to the south-east of the system. The NE-SW flow can unambiguously
be associated with source A2 which might be entering a phase of enhanced activity. The ALMA CO(6-5) observations 
have enabled us to discover a new outflow toward source B with peculiar properties: it is highly asymmetric, resembles
an outburst, is fairly slow (10 km s$^{-1}$), and lacks a jet-like feature along its symmetry axis. In addition, its dynamical 
age is only about 200 years. We argue that these properties result from the extreme youth of source B, which might be 
one of the youngest known protostars. Given the short dynamical age of the structure, significant morphological changes 
are expected to take place over the next few decades. Observing this time evolution would help constrain the formation 
and early evolution of molecular outflows in general, and would have a fundamental impact on our understanding of these 
objects. 

\section*{Acknowledgments}
L.L., L.A.Z., L.F.R.\ and G.P.\ acknowledge the support of DGAPA, UNAM, and of CONACyT (M\'exico). 
LL is indebted to the Alexander von Humboldt Stiftung for financial support. BP is supported by the DFG 
Emmy Noether grant PA1692/1-1. The National Radio 
Astronomy Observatory is a facility of the National Science Foundation operated under cooperative 
agreement by Associated Universities, Inc. This paper makes use of the following ALMA data: 
ADS/JAO.ALMA\#2011.0.00007.SV. ALMA is a partnership of ESO (representing its member states), 
NSF (USA) and NINS (Japan), together with NRC (Canada) and NSC and ASIAA (Taiwan), in cooperation 
with the Republic of Chile. The Joint ALMA Observatory is operated by ESO, AUI/NRAO and NAOJ.

\label{lastpage}

\end{document}